\documentclass[twoside,twocolumn]{article}
\usepackage{slashed, graphicx, amssymb, amsmath, ragged2e,multirow,float,tabu}

\usepackage{blindtext} 
\usepackage{url}
\usepackage[T1]{fontenc} 
\usepackage{microtype} 

\usepackage[english]{babel} 

\usepackage[hmarginratio=1:1,top=32mm,columnsep=20pt,margin=2.0cm]{geometry} 

\usepackage[hang, small,labelfont=bf,up,textfont=it,up]{caption} 
\usepackage{booktabs} 

\usepackage{lettrine} 

\usepackage{enumitem} 
\setlist[itemize]{noitemsep} 

\usepackage{abstract} 

\usepackage{titlesec} 
\renewcommand\thesection{\bfseries\roman{section}} 
\renewcommand\thesubsection{\roman{subsection}} 
\titleformat{\section}[block]{\bfseries\large\scshape\centering}{\thesection.}{0.5em}{} 
\titleformat{\subsection}[block]{\large}{\thesubsection.}{1em}{} 

\usepackage{fancyhdr} 
\usepackage{titling} 

\usepackage{hyperref} 

\usepackage{dblfloatfix}

\usepackage[gen]{eurosym} 

\pretitle{\begin{center}\LARGE\bfseries} 
\posttitle{\end{center}} 
\title{
{\fontfamily{lmss}\selectfont
Lattice Design of a Carbon-Ion Synchrotron based on Double-Bend Achromat Lens}} 
\author{%
\normalsize Xuanhao \textsc{Zhang} \\ 
\normalsize \it{School of Physics, University of Melbourne, Australia} \\ 
\normalsize \it{xuanhao.zhang@unimelb.edu.au} 
}

\date{} 
\renewcommand{\maketitlehookd}{%
\begin{abstract}
A normal-conducting carbon-ion synchrotron design for cancer therapy is proposed. The lattice design is based on the double-bend achromat lens with additional quadrupole magnets. The proposed synchrotron design is compact with a circumference of 55 meters. The lattice is tuned for slow extraction at the third integer resonance and the optical functions are optimised with considerations for reducing the power consumption during operation.
\end{abstract}
}

\begin{document}

\maketitle


\section{introduction}

Irradiation of tumour using carbon-ions is a non-invasive method of cancer therapy. There are several advantages for using using carbon-ions over the more widely available proton therapy, such as greater relative biological effectiveness (RBE) and better depth precision \cite{CarbonTumourTherapy}. However, there is only a handful of carbon-ion therapy facilities around the world partially due to the high cost of building a dedicated facility. The size of the typical synchrotron required to accelerate carbon-ions to therapeutic energies presents a significant cost barrier. In addition, once built the ongoing electricity bill of running the facility could cost more than 1 M$\euro{}$ per year \cite{SEEIIST}. This paper presents a novel synchrotron design for carbon-ion therapy purposes, the design is optimised for both the overall size of the machine and the vertical beam aperture that could potentially reduce the electricity consumption.

\section{design requirements}

The maximum penetration depth required for clinical treatment is 38 cm equivalent depth in water, this corresponds to an energy of 430 MeV/u for carbon ions. The maximum dipole strength is set to 1.5 T and the length of each dipole element should be kept between 2-3 m for most efficient manufacturing costs \cite{Elena}. The momentum compaction factor $\gamma_{t}$ of the machine should be kept above 1.46 to ensure longitudinal beam stability. It is preferable to have long dispersion-free sections in the ring to accomodate RF-cavities and septum magnets to simplify the accelerator cycle.

Design of the synchrotron is constrained by the requirement for a scanning pencil beam treatment method where a smooth beam extraction over the period of 1 s is needed, this is typically achieved through resonance extraction at the third integer tune. The chromaticity of the machine $\xi_{x}$ should be negative for maximum transverse stability of the waiting beam during extraction \cite{Hardt}. Optics of the ring is kept constant during extraction while a sextupole is used to excite the third-order resonance. Particles are smoothly accelerated toward the resonance by either an RF-cavity or betatron core. 
It is desirable to place the resonance sextupole (SXR) in a dispersion-free section of the ring to allow independent resonance and chromaticity control.

The particle motion in phase space in the plane of extraction is given by the Kobayashi Hamiltonian:
\begin{equation}
H = \frac{6 \pi \delta Q}{2} (X^{2}+X'^{2}) + \frac{S_{n}}{4}(3XX'^{2}-X^{3}),
\end{equation}
where $\delta Q$ is the distance to the third order resonance tune $n\pm1/3$, ($X,X'$) are the normalised phase space coordinates and $S_{n}$ is the normalised SXR strength.
Particles leave the stable region along separatrices described by the Kobayashi Hamiltonian and they are extracted from the ring through a combination of electrostatic (ES) and magnetic septa (MS). The size of the stable region in phase space is slowly reduced as particles are extracted to provide a constant beam spill. 
Aligning the separatrices of particles with different momenta at the ES will minimise losses and reduce the size of the septum aperture, this is known as the Hardt condition \cite{Hardt}:
\begin{equation}
D_{n}\cos(\pi - \Delta \mu)+D_{n}'\sin(\pi-\Delta \mu) = -4\pi\xi_{x}/S_{n},
\end{equation}
where $D_{n}, D_{n}'$ is the normalised dispersion function and its derivative at the ES, $\Delta \mu$ is the betatron phase advance between SXR and ES, and $\xi_{x}$ is the chromaticity in the plane of extraction.

\begin{table*}[t]
\begin{center}
\caption{Summary of synchrotron parameters at major operational carbon therapy facilities and the proposed DBA lattice.}
\label{table:SynchSurvey}
\begin{tabular}{llllll|l}
\hline
\hline
                     & PIMMS     & GSI        & Hitachi   & NIRS      & Siemens    & DBA         \\
\hline                                                                
Proton               & $\bullet$ & -          & $\bullet$ & -         & $\bullet$  & -           \\
Carbon	             & $\bullet$ & $\bullet$  & $\bullet$ & $\bullet$ & $\bullet$  & $\bullet$	 \\
No. built            & 2         & 2          & 1         & 5         & 1          & -           \\
Circumference (m)    & 75        & 65         & 60        & 62        & 65         & 55          \\
Cell type            & FODOF     & FD doublet & FODO      & FODO      & FODO       & DBA         \\
No. of cells         & 8         & 6          & 6         & 6         & 6          & 2           \\
No. of dipoles       & 16        & 6          & 12        & 18        & 12         & 12          \\
No. of quads         & 24        & 12         & 12        & 12        & 12         & 14          \\
$E_{max}$ (MeV/u)    & 400       & 430        & 480       & 400       & 430        & 430         \\
Max. $\beta_{y}$ (m) & 15		 & 22		  & 13		  & 13.4 	  & 15.5	   & 11.8		 \\
Tune ($Q_{x}/Q_{y}$) & 1.6666/1.72 & 1.72/1.13 & 1.72/1.43 & 1.672/1.72 & 1.7/1.8 & 1.672/1.72  \\
\hline
\hline
\end{tabular}
\end{center}
\end{table*}

The power consumption of the lattice can be reduced by minimising the size of the vertical beam aperture. The total current of a normal conducting dipole is given by:
\begin{equation}
NI = \frac{B_{gap}h}{\mu_{0}\eta},
\end{equation}
where $B_{gap}$ is the dipole field strength between the poles, $h$ is the pole gap height, $\mu_{0}$ is the vacuum permeability and $\eta\approx1$ for non-saturated dipole with iron yoke. In the case where both beam injection and beam extraction are performed in the radial plane, the vertical chamber aperture is given by \cite{PIMMS}:
\begin{equation}
\textrm{aperture}=\pm\left[n\sqrt{\beta\epsilon/\pi} + \rm{collimator\ margin} \right],
\end{equation}
the first term in the brackets is the vertical betatron envelope for a given RMS emittance, it corresponds to the `good'-field region of the magnets where a good field uniformity is required, the collimator margin corresponds to the `poor'-field region which is typically reserved as a safety margin that is half the size of the `good'-field region. Therefore the current required for normal conducting dipoles is approximately proportional to square-root of the vertical betatron function.

\section{existing designs}

Currently there are 12 facilities around the world that provide cancer therapy treatment using carbon ions, they are located in Japan (6), China (2), Germany (2), Austria (1) and Italy (1). All 12 facilities rely on normal conducting synchrotrons as the main accelerator. The majority of these synchrotron lattice designs can be traced back to one of five projects lead by the following institutions: CERN (PIMMS)\cite{PIMMS}, GSI \cite{GSI}, Hitachi \cite{Hitachi}, NIRS \cite{NIRS} and Siemens \cite{Siemens}. Table \ref{table:SynchSurvey} shows a lattice parameter summary of these five designs.

Most of these designs have separate function, alternating focusing FODO type lattices. The FODO lattice is a relatively simple lattice that offers a high degree of periodicity, however there are several tradeoff factors to the design considerations. There exists an inverse relationship between the size of the main synchrotron and the amplitude of betatron oscillation, this relationship means any lattice design has to carefully balance compactness of the machine and beam parameters. 
In addition, the basic FODO lattice does not have any dispersion free sections and often requires three or more families of quadrupoles to tune the lattice for slow extraction. The basic FODO lattice can be modified to accomodate dispersion free drifts but at the cost of additional quadrupoles and overall size. In the case of the PIMMS design, the number of quadrupoles magnets in the main ring is double the number in that of a typical FODO lattice.

\section{lattice design}

The motivation for using the Double-Bend Achromat (DBA) \cite{DBA} is mainly driven by the preference for dispersion free drift sections. In comparison with a compact PIMMS type FODOF lattice \cite{KoreanPIMMS}, the DBA requires fewer number of quadrupole magnets and independent power supplies.

\begin{figure}[h]
	\centering
	\caption{Original double-bend achromat arc consist of two dipoles and five quadrupoles arranged symmetrically. \cite{DBA}}
	\includegraphics[width=1.0\linewidth]{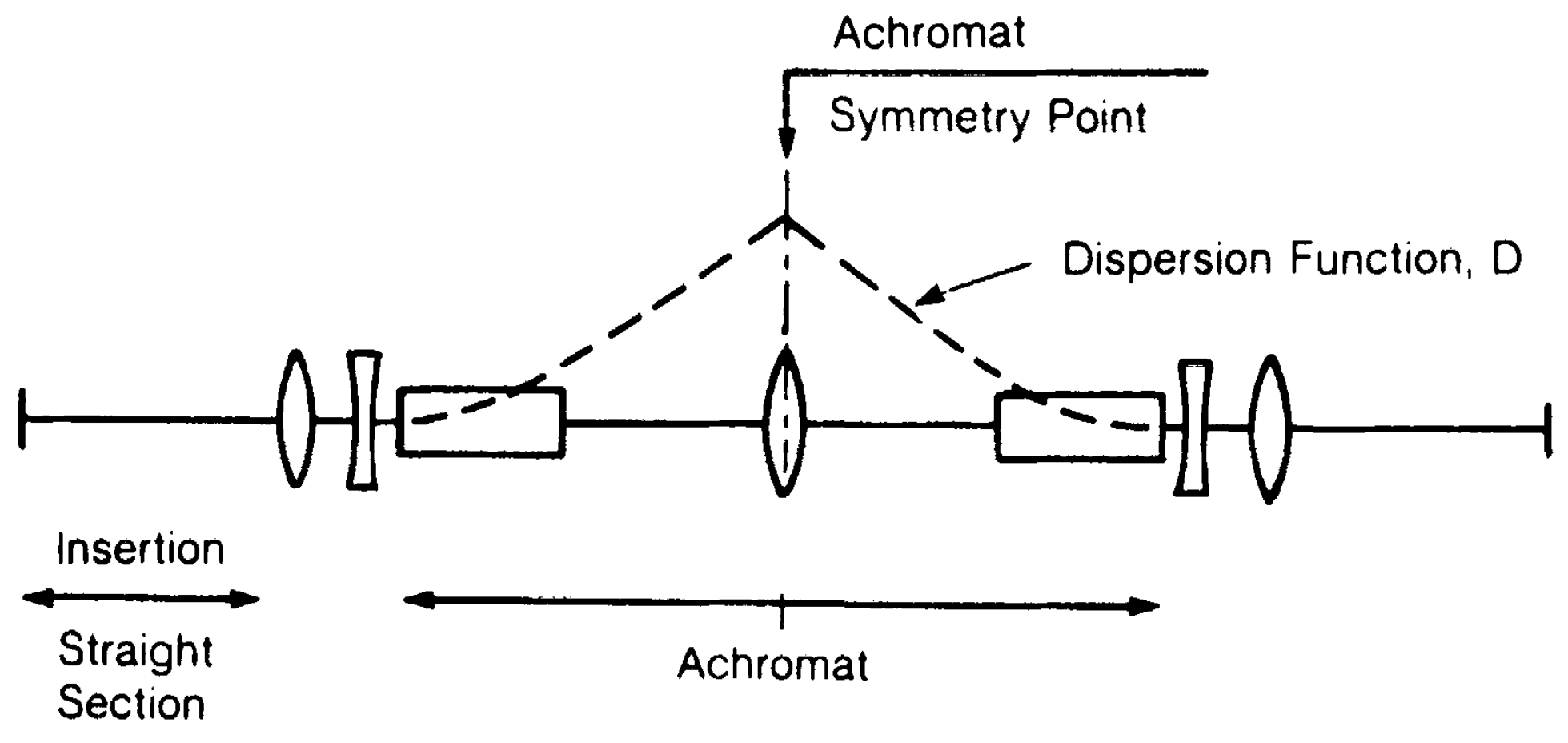}
	\label{fig:OriginalDBA}
\end{figure}

Figure \ref{fig:OriginalDBA} shows the layout of the original DBA. The DBA was initially designed as a candidate for electron storage rings where small beam divergences and very low emittance in long drift sections are required to accomodate insertion devices. It was shown that the original DBA can be optimised for minimum betatron function in the long drifts by changing the distance between the bending dipoles and the central quadrupole \cite{Sommer} as illustrated in Figure \ref{fig:OriginalDBA}. This property is exploited in the proposed DBA design in this paper to minimise the vertical betatron envelope. 

\begin{figure}[h]
	\centering
	\caption{Betatron amplitude and dispersion functions of the proposed DBA lattice at $Q_{x} = 1.672, Q_{y}=1.72$.}
	\includegraphics[width=1.0\linewidth]{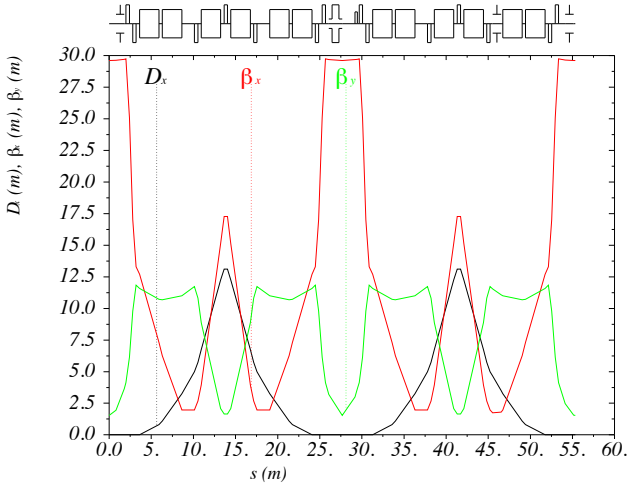}
	\label{fig:madxplot}
\end{figure}

The lattice design was performed using MADX \cite{MADX}.
First, two 180$^{\circ}$ DBA were joined together to form a closed ring with two dispersion-free long drifts. The total bending length is adjusted to 27.72 m corresponding to a maximum dipole field of 1.5 T.
Each dipole section was subdivided into three dipoles that are each 2.31 m in length. An additional quadrupole was added symmetrically on both sides of the central quadrupole to achieve a tune that is close to the third integer. It was necessary to shift the additional quadrupoles along the ring by one dipole element to provide the required phase advance between the SXR, ES and MS for slow extraction scheme. 
The SXR and a placeholder RF-cavity is located at one of the dispersion-free section, while both injection and extraction magnetic septa are located at the other dispersion-free section. The choice of septa arrangement potentially limits the orientation of injection and extraction to opposite sides of the ring. It is possible to swap the injection septum with the RF cavity by increasing the length of the dispersion-free drift length. 
The ES is placed upstream of the extraction MS that satisfies the Hardt condition, where $D_{n} = 2.60$, $D_{n}' = -2.44$, $\Delta\mu(SXR,ES) = 226^{\circ}$ and $\Delta\mu(ES,MS) = 70.5^{\circ}$.
Space is reserved in dispersive regions along the ring for a pair of chromaticity sextupoles. As such, the chromaticity of the machine and the resonance conditions can be controlled separately.
Additional work is required to include bump magnets for injection and extraction.

\begin{figure}[h]
	\centering
	\caption{Schematics of the synchrotron, the designed dipole magnets are 30$^{\circ}$ sector dipoles. QF: focusing quadrupole, QD: de-focusing quadrupole.}
	\includegraphics[width=1.0\linewidth]{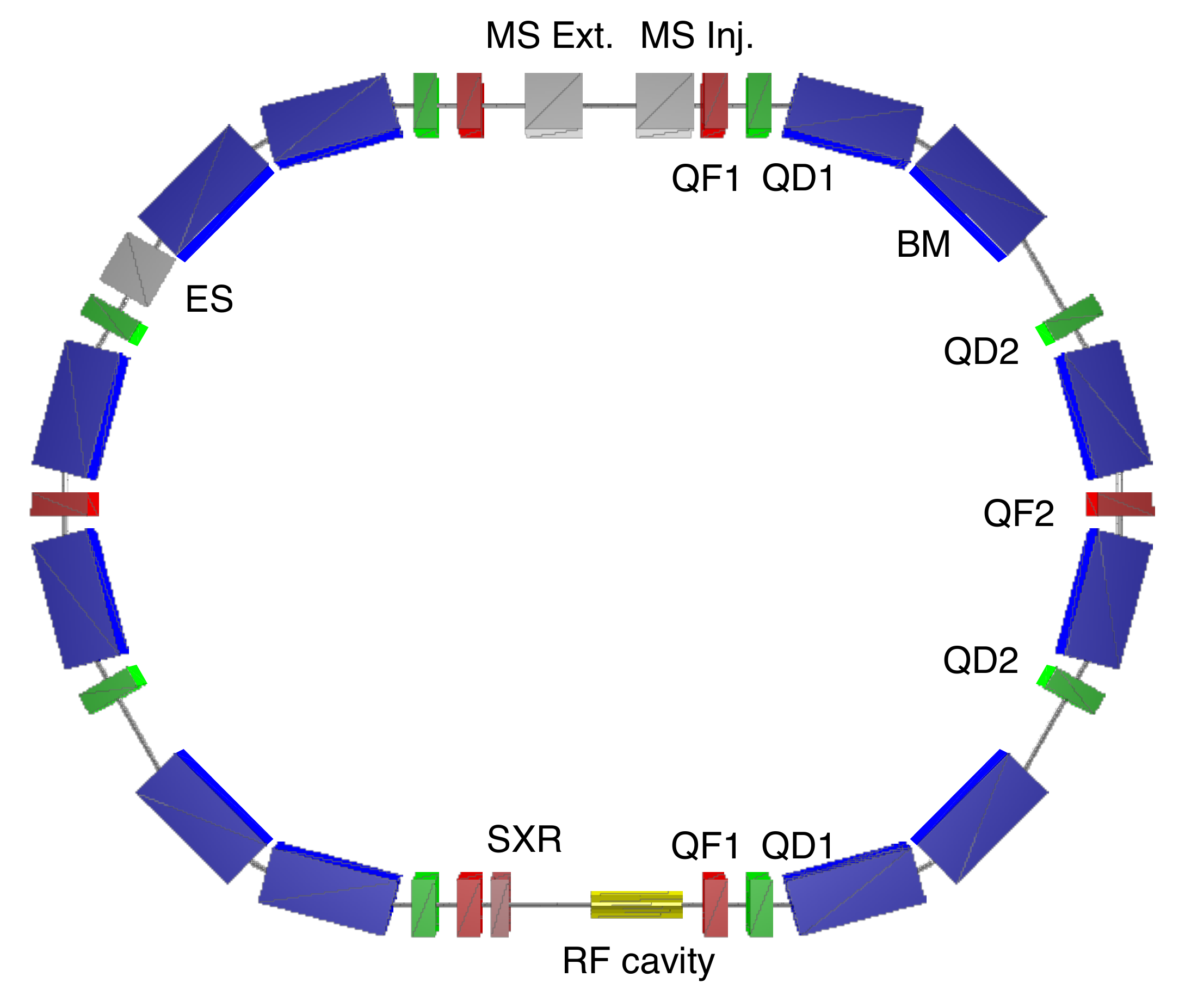}
	\label{fig:layout}
\end{figure}

The vertical betatron envelope was optimised by varying the drift lengths and quadrupole strengths. The minimum drift length between any element was limited arbitrarily to 0.2 m as an engineering margin. Figures \ref{fig:madxplot} and \ref{fig:layout} show the lattice functions and the schematic layout of the final optimised design. Specifications of the proposed DBA lattice is included in Table \ref{table:SynchSurvey} for comparison with exisiting designs. Detailed parameters of the final design can be found in Table \ref{table:detailed}.

\begin{table}[h]
\centering
\caption{Detailed parameters of the proposed DBA lattice.}
\begin{tabular}{l l}
\hline
\hline
Transition gamma, $\gamma_{t}$ & 1.742\\
Magnetic rigidity, $\rho B$ & 6.624 T-m\\
Maximum $\beta_{x},\beta_{y}$ & 29.745, 11.839 m\\
Maximum dispersion, $D_{x}$ & 13.109 m\\
Natural chromaticities, $\xi_{x}$, $\xi_{y}$ & -1.093, -1.276\\
Dipole length & 2.31 m\\
Dipole strength & 1.5 T\\
Quadrupole length & 0.4 m\\
QF1 strength & 1.049 m$^{-2}$\\
QF2 strength & 0.800 m$^{-2}$\\
QD1 strength & -1.299 m$^{-2}$\\
QD2 strength & -0.650 m$^{-2}$\\ 
\hline
\hline
\end{tabular}
\label{table:detailed}
\end{table}

Typically, ES is placed in a region with relatively large horizontal betatron amplitude to enhance the `kick' from the electric field and prevent the extracted particles from colliding with the MS wire. 
The horizontal displacement at the MS, $\Delta x_{MS}$, due to a deflection of $\theta_{ES,x}$ from the ES upstream is given by:
\[
\Delta x_{MS} = \theta_{ES,x}\sqrt{\beta_{ES,x}\cdot\beta_{MS,x}}\cdot \sin(\Delta\mu),
\]
where $\Delta\mu$ is the betatron phase difference between MS and ES, ($\beta_{ES,x},\beta_{MS,x}$) are the horizontal betatron functions at ES and MS respectively.
The deflection $\theta_{ES,x}$ of a charged particle beam in an electric field is given by:
\[
\theta_{ES,x} = \tan^{-1}\left[ \frac{E_{x}\ l_{ES}\ q}{p\ \beta \times 10^{9}}\right],
\]
where $E_{x}$ is the electric field strength, $l_{ES}$ is the effective field length of ES, $p$ is the momentum in GeV/c and $\beta$ is the relativistic velocity.
In the current design, an electric field of 61.1 kV/cm will result in a 0.6 mm displacement at maximum extraction energy. The field required is within feasible limits of ES field strength and the resultant deflection is enough to clear a `thin' MS similar to that used in the PIMMS design.

\section{conclusions}

A normal conducting synchrotron was designed to reduce the size of carbon-ion therapy facilities and potentially lessen the electricity cost of operating the machine. The lattice has been designed to satisfy the Hardt condition for third-order resonant extraction.


\section*{acknowledgement}

This work is supported by the Australian Government Research Training Program Scholarship.

\renewcommand\refname{references}
\bibliographystyle{./bibstyles/h-physrev}
\bibliography{Bib}


\end{document}